\documentclass[12pt,a4paper]{article}
\usepackage{amssymb,amsmath,amscd,epsfig}
\title{Neutrino oscillations: deriving the plane-wave approximation in the wave-packet approach.}
\author{\copyright~
 O.V. Lychkovskiy
\thanks{e-mail: lychkovskiy@itep.ru}~;
\\ {\small\it Institute for Theoretical and Experimental Physics, Moscow, Russia;}\\
{\small\it Moscow Institute of Physics and Technology, Dolgoprudny, Russia.}
}

\begin{document}

\newcommand{\nue}{\nu_e}
\newcommand{\nul}{\nu_l}
\newcommand{\antinue}{\bar{\nu}_e}
\newcommand{\ar}{\rightarrow}
\newcommand{\C}{{\rm C}}
\newcommand{\Hy}{{\rm H}}
\newcommand{\F}{{\rm F}}
\newcommand{\Co}{{\rm Co}}
\newcommand{\N}{{\rm N}}
\newcommand{\Ox}{{\rm O}}
\newcommand{\Fe}{{\rm Fe }}
\newcommand{\be}{\begin{equation}}
\newcommand{\ee}{\end{equation}}
\newcommand{\bc}{\begin{cases}}
\newcommand{\ec}{\end{cases}}
\newcommand{\B}{{\bf B}}
\newcommand{\Bp}{{\bf B_\perp}}
\newcommand{\rb}{\right)}
\newcommand{\lb}{\left(}
\newcommand{\kpc}{{\rm kpc}}
\date{}

\maketitle

\begin{abstract}

{The plane-wave approximation is widely used in the practical calculations concerning neutrino oscillations.
A simple derivation of this approximation starting from the neutrino wave-packet framework is presented.}

\end{abstract}

The purpose of the present work is to give a simple derivation of the plane-wave approximation, which is widely used in the neutrino oscillation theory, starting from wave packets. This subtle issue is frequently omitted in the reviews and lectures concerning neutrino oscillations, although it is rather important, at least from the pedagogical point of view.
The present work is a piece of lectures given at the 36th ITEP winter school of physics. Other topics discussed in the lectures were common: three-flavor neutrino mixing and oscillations, MSW-effect, neutrino oscillation experiments.  These topics are widely discussed and described in the literature. The review of these topics may be found elsewhere\cite{in preparation}.

Neutrinos are neutral leptons with very low (less than $\sim 1$eV) mass. They enter the Standard Model as the neutral counterparts of the three charged leptons: electron $e$, muon $\mu$ and tau-lepton $\tau$.
Accordingly, there are three types of neutrinos in the Standard Model: electron neutrino $\nu_e,$ muon neutrino $\nu_\mu$ and tau-neutrino $\nu_\tau$ (as in the case of charged leptons, this three types are called "flavors"). However, for our purposes it is reasonable to consider the two-flavor model: it has the advantage of clarity, and all principal points may be studied within it. Moreover in many cases one can use the two-flavor scheme as an accurate approximation to the three-flavor oscillations.

As an example of the two-flavor model we consider the Standard Model without tau-lepton and tau-neutrino.\footnote{To avoid anomalies one should also exclude the third generation of quarks.} It contains four leptons: $e, ~\mu, ~\nu_e$ and $\nu_\mu.$

Oscillation phenomenon is based on the fact that neutrino of a definite flavor (i.e. flavor eigenstate) does not possess a definite mass (i.e. does not coincide with a mass eigenstate). Instead flavor eigenstates are quantum mechanical superpositions of mass eigenstates $\nu_1$ and $\nu_2$:

\begin{equation}
\begin{tabular}{l}
$\nu_e = c \nu_1 + s \nu_2,$
\\ $\nu_\mu = - s \nu_1 + c \nu_2,$
\end{tabular}
\end{equation}
where $c$ and $s$ denote cosine and sine of the mixing angle $\theta$:

\begin{equation}
c \equiv \cos\theta, ~~~s \equiv \sin\theta.
\end{equation}

The mass basis set is orthonormal:

\begin{equation}\label{orthonormallity of mass eigestates}
\begin{tabular}{l}
$\langle\nu_1|\nu_1\rangle =\langle\nu_2|\nu_2\rangle =1,$
\\
$\langle\nu_1|\nu_2\rangle =0.$
\end{tabular}
\end{equation}
As $c^2+s^2=1,$ flavor basis set is also orthonormal:
\begin{equation}\label{orthonormallity of flavor eigestates}
\begin{tabular}{l}
$\langle\nu_e|\nu_e\rangle =\langle\nu_{\mu}|\nu_{\mu}\rangle =1,$
\\
$\langle\nu_e|\nu_{\mu}\rangle =0.$
\end{tabular}
\end{equation}

The interaction part of the Standard Model Lagrangian, diagonal in the flavor basis, ensures that neutrinos are created (and detected) always in flavor eigenstates.
For definiteness, we consider the $\nu_e$ creation on the proton through the charge current:
$$e+p\rightarrow n+\nu_e.$$
Wave function of the just created ($t=0$) electron neutrino reads
\begin{equation}\label{wave function at t=0}
  \Psi(x,0)= F(x)(e^{i p_1 x}c \nu_1+s e^{i p_2 x}\nu_2),
\end{equation}
where
\begin{equation}\label{expansion for p}
  p_i=\sqrt{E^2-m_i^2}\simeq E-\frac{m_i^2}{2 E}, ~~~i=1,2.
\end{equation}
 Here $F(x)$ is an envelope of the neutrino wave packet at $t=0$, $E$ is neutrino energy, $p_i$ is momentum of the $i$-th neutrino eigenstate.\footnote{There is a wide discussion in the literature, whether one should ascribe different momenta and equal energies to different mass eigenstates, or use some other prescription, say, different energies and equal momenta. The self-consistent way to resolve this ambiguity is to consider the specific neutrino creation process in the QFT framework. Such studies show that equal energy prescription is a good approximation in practical cases \cite{Dolgov:2005nb,Dolgov:2005vj}.}
As typical neutrino energies are always orders of magnitude greater than 1 eV, neutrinos are ultra-relativistic: $\frac{m_i^2}{2 E^2}\ll 1.$ For simplicity we consider one-dimensional motion. Generalization on three dimensions is straightforward.

The envelope $F(x)$ is peaked at $x=0$ and vanishes for  $|x|\gtrsim l/2,$ where $l$ is the size of the neutrino wave packet. The normalization condition is
\begin{equation}\label{F norm}
 \int dx |F(x)|^2=1.
\end{equation}

Specific form of $F(x)$ depends on the details of the neutrino creation process. It is shown below that the oscillation pattern is insensitive to this form, provided inequalities (\ref{condition 1}) and (\ref{condition 2}) hold.

In order for the wave function $\Psi(x,0)$ to describe indeed an {\it electron} neutrino, one should demand

\begin{equation}
\label{condition 1}
  l (p_2-p_1)\simeq l\frac{m_2^2-m_1^2}{2 E}\ll 1.
\end{equation}
This allows one to rewrite Eq.(\ref{wave function at t=0}) in the following form:
\begin{equation}
\label{wave function at t=0, approximation}
  \Psi(x,0)\simeq e^{i p_1 x} F(x)(c \nu_1+s \nu_2)=e^{i p_1 x} F(x)
  \nu_e.
\end{equation}
One can see that the neutrino wave function at $t=0$ represents, as expected, a state with definite flavor - namely a state of electron neutrino. Furthermore, it follows from (\ref{orthonormallity of flavor
eigestates}), (\ref{F norm}) and (\ref{wave function at t=0, approximation}) that $\Psi(x,0)$ is normalized to unity.

The next step is to consider time evolution of the neutrino wave function. It is known from quantum mechanics that the wave packet time evolution results in the appearance of the
phase $e^{-iEt}$ and in the
 motion of the wave packet with the speed $v\equiv \partial E/\partial p =  p/E.$\footnote{The third effect of evolution, the spreading of the wave packet, is negligible for ultra-relativistic particles.}
In our case this leads to
\begin{equation}\label{wave function}
  \Psi(x,t)= e^{-i E t }(F(x-v_1 t) e^{i p_1 x} c \nu_1 + F(x-v_2 t) e^{i p_2 x} s
  \nu_2),
\end{equation}
where
\begin{equation}
v_i\equiv p_i/E \simeq 1- \frac{m_i^2}{2 E^2}.
\end{equation}
Velocities $v_1,~v_2$ are very close to each other and to unity. For the time being we take $v_1=v_2=v$ and $v=1.$
The validity of the first assumption is discussed below. As for the second one, it is a technical one and is admitted only for the simplicity of notations.
Let us modify Eq.(\ref{wave function}) with the use of
Eq.(\ref{expansion for p}):
\begin{equation}\label{wave function, approximation}
  \Psi(x,t)\simeq e^{-i E t + i p_2 x} F(x-t) \left(\exp({i \frac{m_2^2-m_1^2}{2E}}x) c \nu_1 + s
  \nu_2\right).
\end{equation}

Note that, due to the factor $F(x-t),$  $\Psi(x,t)$ vanishes everywhere but the $l/2$-vicinity of the point $x=t.$

Assume that at $t=L$ one measures the neutrino flavor (this implies automatically that one finds neutrino coordinate to be in the vicinity of $x=L$). In practice to accomplish such a measurement one needs to register some neutrino-induced reaction in the detector, which is sensitive to the neutrino flavor. The probability to observe electron neutrino,
$P(\nu_e\rightarrow\nu_e),$ reads
$$P(\nu_e\rightarrow\nu_e)=|\langle\nu_{e}|\Psi(x,L)\rangle |^2\simeq$$
\begin{equation}\label{P(nu_e-->nu_e)}
\int dx |F(x-L)|^2~\left|\langle c \nu_1+s \nu_2|\exp({i
\frac{m_2^2-m_1^2}{2E}}L) c \nu_1 + s
  \nu_2\rangle \right|^2=1-\sin^2 2\theta \sin^2 \frac{\Delta m^2 L}{4E}, ~~~~~
\end{equation}
$$\Delta
m^2\equiv m_2^2-m_1^2.$$

Analogously,

\begin{equation}\label{P(nu_e-->nu_mu)}
P(\nu_e\rightarrow\nu_{\mu})=\sin^2 2\theta \sin^2 \frac{\Delta
m^2 L}{4E}.
\end{equation}

Thus propagating neutrino changes its flavor. It is this phenomenon which is called neutrino oscillations. The oscillation probability is a periodic function with the period

$$L^{osc}\equiv \frac{4\pi E}{\Delta
m^2 }.$$

Note that the neutrino wave packet envelope $F$ enters the expression for the probability of oscillations, Eqs.(\ref{P(nu_e-->nu_e)}), only through the normalization integral $\int dx |F(x-L)|^2,$ which is equal to unity according to the normalization condition (\ref{F norm}). Thus the final result does not depend on $F.$
Therefore we could formally take $F$ equal to 1 from the very beginning, simultaneously throwing away the normalization integral. This trick is the essence of the commonly used plane-wave approximation, in which coordinate parts of neutrino wave functions are taken to be plane waves:

\begin{equation}\label{plane wave}
  \Psi(x,t)= e^{-i E t } (e^{i p_1 x} c \nu_1 +  e^{i p_2 x} s
  \nu_2)
\end{equation}
This wave function is an energy eigenstate, which guarantees that at $x=0$ (neutrino source coordinate) and at arbitrary $t$
it describes electron neutrino\cite{Vysotsky}:
\begin{equation}\label{}
  \Psi(0,t)= e^{-i E t } \nu_e.
\end{equation}
Furthermore, one can from the very beginning omit the unessential common phase factor $e^{-iEt}.$ In the plane-wave approximation this factor is the only place where time $t$ appears.

 The probability to observe a muon neutrino at the distance $L$ from the source is calculated as follows:
$$P(\nu_e\rightarrow\nu_{\mu})=|\langle\nu_{\mu}|\Psi(L)\rangle |^2=$$
\begin{equation}\label{P(nu_e-->nu_mu) in plane wave approx}
|\langle -s \nu_1+c \nu_2|e^{i p_1 L} c \nu_1 + e^{i p_2 L} s
  \nu_2\rangle |^2=\sin^2 2\theta \sin^2 \frac{\Delta m^2 L}{4E},
\end{equation}
which coincides with the result obtained in the wave-packet formalism.

Now let us discuss when the approximation $v_1=v_2$ is valid.
Evidently, this is the case as long as the spatial separation of the wave packets is insignificant:
\be
F(x-v_1 t)\simeq F(x-v_2 t),
\ee
which gives
\be
|v_2-v_1|L \ll l.
\ee

The later inequality may be
rewritten as
\be
\label{condition 2}
2\pi L/L^{osc}\ll lE.
\ee
This is always valid for neutrinos which are created and registered at the Earth. If we deal with neutrinos of cosmic origin (e.g. supernova neutrinos), this condition may fail. However, for such neutrinos matter effect (MSW-effect), which was not discussed here, normally plays a crucial role.
\newline
\newline
{\bf\Large Acknowledgments}\\
The author is grateful to L.B.Okun, M.I.Vysotsky, M.V.Rotaev and A.A.Mamonov for valuable discussions. The
work was financially supported by the Dynasty Foundation scholarship, RF
President grant NSh-4568.2008.2, RFBR grants 07-02-00830-a and RFBR-08-02-00494-a.

\end{document}